# A Cu²⁺ (S = ½) Kagomé Antiferromagnet: Mg$_x$Cu$_{4-x}$(OH)$_6$Cl$_2$

Shaoyan Chu‡, Tyrel M. McQueen†, Robin Chisnell§, Danna E. Freedman†, Peter Müller†, Young S. Lee§ and Daniel G. Nocera*,†

†Department of Chemistry, §Department of Physics, and the ‡Center for Materials Science and Engineering,
Massachusetts Institute of Technology, 77 Massachusetts Avenue, Cambridge, MA 02139 USA

Spin-frustrated systems are one avenue to induce macroscopic quantum states in materials. However, experimental realization of this goal has been difficult owing to the lack of simple materials and, if available, the separation of unusual magnetic properties arising from exotic magnetic states from behavior associated with chemical disorder, such as site mixing. Here we report the synthesis and magnetic properties of a new series of magnetically frustrated material, Mg$_x$Cu$_{4-x}$(OH)$_6$Cl$_2$. Owing to the substantially different ligand-field chemistry of Mg²⁺ and Cu²⁺, site disorder within the kagomé layers is minimized, as directly measured by x-ray diffraction. Our results reveal that many of the properties in this material and related systems are not due to disorder of the magnetic lattice, but rather reflect an unusual ground state.

Geometric frustration of magnetic ordering on triangle-based lattices is thought to be one avenue to inducing macroscopic quantum states in electron systems.[1] Due to the triangular arrangement of ions, it is impossible to satisfy all nearest-neighbor interactions simultaneously (Fig. 1a). This 'frustration' suppresses classical magnetic long-range order (LRO) and is thought to be capable of resulting in novel quantum states such as the resonating-valence-bond (RVB) or 'spin-liquid' ground state for a two-dimensional (2D) S = ½ antiferromagnet.[2] However, 'structurally perfect' frustrated materials are rare; frequently, triangular lattices undergo a structural distortion at low temperature, relieving the magnetic frustration and giving rise to a classical ground state.[3-6] One of the few known examples is the x = 1 end-member of the paratacamite series Zn$_x$Cu$_{4-x}$(OH)$_6$Cl$_2$.[7-10] It has a perfect 2D kagomé (corner-sharing triangle) lattice of Cu²⁺ (S = ½) ions in Jahn-Teller distorted O$_4$Cl$_2$ octahedra, separated by layers of Zn²⁺ in O$_6$ octahedra. However, the chemical similarity between Zn²⁺ and Cu²⁺ combined with the difficulty in differentiating Zn or Cu by x-ray and neutron diffraction techniques has complicated studies of this material, as site mixing of Zn²⁺ and Cu²⁺ in the kagomé planes may also account for the observed behaviors.[11] Herein we report the structural and magnetic characterization of a new series of compounds, Mg$_x$Cu$_{4-x}$(OH)$_6$Cl$_2$, which is isostructural with paratacamite.[12] Whereas both Mg and Cu can occupy the interplane O$_6$ site, the ligand-field chemistry of non Jahn-Teller active Mg strongly disfavors its residency within the tetragonally-distorted O$_4$Cl$_2$ coordination sites in the kagomé layers. This disparity in ligand-field chemistry of Mg and Cu ensures minimal substitution of Cu by Mg into the kagomé interlayer.

Synthesis of Mg$_x$Cu$_{4-x}$(OH)$_6$Cl$_2$ proceeds in an analogous manner to paratacamite.[8] In a typical reaction, Cu$_2$(OH)$_2$CO$_3$ and a large excess of MgCl$_2$·6H$_2$O (2-4 Mg : 1 Cu) are combined at 130-190 °C under hydrothermal conditions. After 2-3 days, a blue-green powder of Mg$_x$Cu$_{4-x}$(OH)$_6$Cl$_2$ forms at the base of the reaction vessel. The Mg content in the product was controllable, with higher Mg excesses and lower temperatures producing samples with larger x. Polycrystalline samples with x = 0.39 (**1**), 0.54 (**2**), and 0.75 (**3**) are reported here. Attempts to prepare samples with higher x were unsuccessful. Millimeter-size crystals with x = 0.33 (**4**), 0.65 (**5**), and 0.75 (**6**) were grown under hydrothermal conditions in a temperature gradient from powders placed at the hot end (190 °C) of the reaction vessel. Blue-green, octahedral crystals formed at the cold end. Detailed synthesis procedures and crystallographic x-ray data can be found in the SI.

As shown in Fig. 1b, the trigonal crystal structure of Mg$_x$Cu$_{4-x}$(OH)$_6$Cl$_2$ consists of 2D kagomé layers perpendicular to the c axis. These layers are built from corner-sharing CuO$_4$ plaquettes, which are tilted with respect to each layer. Triangles of the networks are bridged by MgO$_6$ octahedra between the layers separated by Cl⁻ anions. Nominally, the in-plane, Jahn-Teller distorted, O$_4$Cl$_2$ sites are entirely occupied by Cu²⁺, with Mg²⁺ being incorporated solely into the interplane O$_6$ site, and thus the formula can be logically written as (Mg$_x$Cu$_{1-x}$)Cu$_3$(OH)$_6$Cl$_2$. To quantify the maximum amount of Mg on in-plane sites, several different tests were performed based on the single crystal data. First, refinements were performed assuming no mixing of Mg into the kagomé planes. Subsequently, the Mg:Cu ratio in the plane was allowed to vary, adding one additional parameter to the refinement. By the Hamilton R-ratio test,[13] including this one extra parameter is barely on the edge of statistical significance at the 95% confidence level (1.008, 1.008, and 1.008 for **4**, **5**, and **6** respectively, versus a cutoff of 1.008). Furthermore, the freely refined Mg content on the kagomé planes is small in each case, at most 5.5 standard deviations away from zero (0.005(13), 0.032(9), and 0.047(9) for **4**, **5**, and **6**, respectively). Thus the amount of mixing, if any, is small. As a more robust quantification of the maximum amount of Mg on the in-plane sites, Fig. 1c shows R1, obtained from refinements of single crystal x-ray data for **4**, **5**, and **6** at 100 K, at various fixed levels of Mg in the kagomé layers (normalized to the value obtained with no mixing). The minimum is sharp in each case, and consistent with at most 3% of Cu²⁺ being replaced by Mg²⁺. This is significantly less than the 7-10%[11] that has been proposed for $Zn_xCu_{4-x}(OH)_6Cl_2$ on the basis of neutron diffraction data, and confirms that, at least in $Mg_xCu_{4-x}(OH)_6Cl_2$, the observed magnetic behavior cannot be due to significant disorder within the 2D kagomé planes.

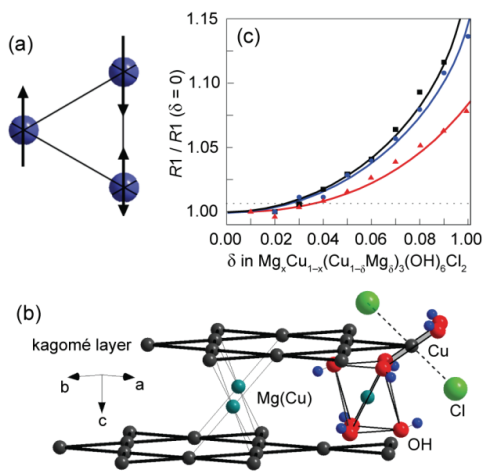

**Figure 1.** (a) The simplest geometric unit on which magnetic frustration can occur is a triangle. (b) Idealized structure of Mg$_x$Cu$_{4-x}$(OH)$_6$Cl$_2$, illustrating the 2D kagomé arrangement of Cu²⁺ ions separated by interlayer cations. (c) A plot of the x-ray refinement statistic R1 as a function of δ, which is the fraction of the sites in the kagomé plane that are occupied by Mg for a fixed x content, Mg$_x$Cu$_{1-x}$(Cu$_{1-δ}$Mg$_δ$)$_3$(OH)$_6$Cl$_2$ (x = 0.33 (□, black), x = 0.65 (○, blue), x = 0.75 (Δ, red)). R1 is normalized to the value when δ = 0; at most, only 3% of the in-plane Cu²⁺ atoms are replaced by non-magnetic Mg²⁺. The dashed line corresponds to the 95% confidence level for one extra parameter in the Hamilton R-ratio test. The lines are simply provided to guide the eye.

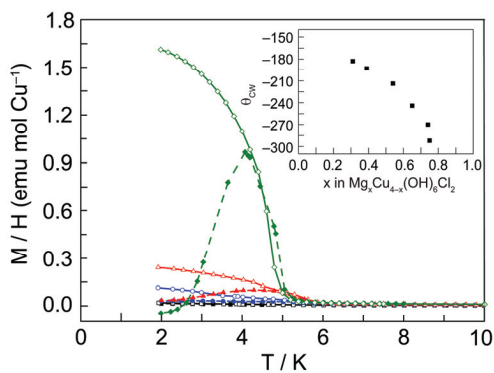

**Figure 2.** Field cooled (open symbols) and zero-field cooled (filled symbols) magnetization data ($H_{appl}$ = 200-500 Oe) of $Mg_xCu_{4-x}(OH)_6Cl_2$ samples (x = 0.75 (□, black), x = 0.54 (○, blue), x = 0.39 (△, red), x = 0.33 (◇, green)). Inset: the Curie-Weiss temperature for samples with varying x extracted from fits of the high temperature magnetic susceptibility.

Field cooled (FC) and zero-field cooled (ZFC) magnetization data were collected for samples **1-4**, $Mg_xCu_{4-x}(OH)_6Cl_2$ (x = 0.33, 0.39, 0.54, and 0.75), at 2-10 K under applied dc fields of 200-500 Oe (Fig. 2). For all samples, the estimated susceptibility, M/H, is small from room temperature down to 5-6 K. Below 5-6 K, M/H increases sharply, indicative of ferromagnetic-like magnetic ordering (likely canted antiferromagnetism). Although the magnitude of the transition ∼ 5 K changes significantly with composition, the temperature at which the transition occurred was essentially composition independent (Figure S6 in the Supporting Information). With increasing substitution of Mg on the interlayer site, the maximum value of the susceptibility decreases, with a corresponding decrease in the splitting between the FC and ZFC data. By x = 0.75, only a faint upturn remains.

The suppression of ordering as x increases is explained if the ferromagnetic-like behavior arises from magnetic coupling between the kagomé layer $Cu^{2+}$ and the interlayer $Cu^{2+}$ ions. When x is small, there is significant coupling between neighboring kagomé layers, giving rise to regions of magnetic order, as is observed in the x = 0 end member, clinoatacamite.[14] As the number of interlayer $Cu^{2+}$ ions is reduced (x increases), there are fewer magnetic exchange pathways between layers. This leads to a progressive decrease in the size and number the magnetically ordered domains, which exist only where interlayer coupling is present. Consequently, as x increases, the sample displays a smaller ferromagnetic response. This model also explains why the temperature at which the upturn appears is composition-independent. While the number of interlayer $Cu^{2+}$ ions affects the fraction of the sample exhibiting ferromagnetism, the intrinsic temperature at which ordering occurs it set by the magnitude of the coupling through a given interlayer $Cu^{2+}$. The size of this coupling is determined by the Cu–O bond lengths and Cu–O–Cu bond angles. The x-ray data show that the structural parameters depend only weakly on composition (Fig. S4), and thus the temperature at which ferromagnetism appears should only be weakly x-dependent.

Interlayer $Cu^{2+}$ magnetic coupling is further supported by the change in the Curie-Weiss temperature ($\theta_{CW}$), which is a measure of the strength of the magnetic interactions, across the series. The Curie-Weiss constants for these samples were extracted from a fit of the high temperature inverse susceptibility data (Fig. S5). The inset of Fig. 2 shows $\theta_{CW}$ as a function of x. The values are large and negative, indicating strong antiferromagnetic exchange. The degree of magnetic frustration can be ascertained from the ratio of the ordering temperature to the Curie-Weiss constant (here $|\theta_{CW}|$ / $T_N \geq 30$, with a minimum cutoff of 10 taken to mean strong frustration). As x approaches 0.75, $\theta_{CW}$ becomes more negative by nearly a factor of two, suggesting a strong increase in antiferromagnetic interaction and magnetic frustration. Inasmuch as the actual geometry is unchanged over the entire series (e.g. in-plane Cu–O–Cu angle is 119.12(12)° in x = 0.33 and 119.06(9)° in x = 0.75), the strength of magnetic coupling within the planes is constant. Thus the interactions between in-plane and interlayer sites, due to $Cu^{2+}$ in the $O_6$ octahedra, are ferromagnetic, and give rise to the magnetic order observed. These data also suggests that as x approaches unity, no magnetic transition will remain despite only a maximum of a small amount of Mg/Cu site disorder. The absence of magnetic disorder is similar to what has been observed in $Zn_xCu_{4-x}(OH)_6Cl_2$[8], and is consistent with an exotic ground state in these materials.

Interlayer $Cu^{2+}$ atoms in $Mg_xCu_{4-x}(OH)_6Cl_2$ exhibit ferromagnetic coupling to in-plane, kagomé $Cu^{2+}$ ions. However, magnetic ordering is suppressed when $Cu^{2+}$ ions are absent from the interlayer. Within the kagomé layers, minimal substitution of $Mg^{2+}$ for $Cu^{2+}$ is observed (≤3%) owing to the significantly different ligand-field chemistry of these two ions. The absence of magnetic order and minimal site disorder within the kagomé planes suggests an unconventional magnetic ground state for $Mg_xCu_{4-x}(OH)_6Cl_2$. These results imply that the lack of a magnetic ordering transition in materials with this structure type, such as $ZnCu_3(OH)_6Cl_2$, which also does not magnetically order down to temperatures of 50 mK,[10,11] is not due to chemical disorder but is indeed a result of the high spin frustration within the kagomé planes.

**Acknowledgements.** This work was supported by the MRSEC Program of the NSF under award number DMR-0819762 and DOE under Grant No. DE-FG02-04ER46134. The authors acknowledge helpful discussions with Oleg Tchernyshyov, Yiwen Chu and Patrick Lee.

**References**

(1) Moessner, R.; Ramirez, A. R. *Physics Today* **2006**, *59*, 24-29.
(2) Anderson, P. W. *Mat. Res. Bull.* **1973**, *8*, 153-160.
(3) Greedan, J. E. *J. Mat. Chem.* **2001**, *11*, 37-53.
(4) Ramirez, A. P. *Ann. Rev. Mat. Sci.* **1994**, *24*, 453-480.
(5) McQueen, T. M.; Stephens, P. W.; Huang, Q.; Klimczuk, T.; Ronning, F.; Cava, R. J. *Phys. Rev. Lett.* **2008**, *101*, 166402.
(6) Collins, M. F.; Petrenko, O. A. *Can. J. Phys.* **1997**, *75*, 605-655.
(7) Braithwaite, R. S. W.; Mereiter, K.; Paar, W. H.; Clark, A. M. *Min. Mag.* **2004**, *68*, 527-539.
(8) Shores, M. P.; Nytko, E. A.; Bartlett, B. M.; Nocera, D. G. *J. Am. Chem. Soc.* **2005**, *127*, 13462-13463.
(9) Lee, S. H.; Kikuchi, H.; Qiu, Y.; Lake, B.; Huang, Q.; Habicht, K.; Kiefer, K. *Nat. Mat.* **2007**, *6*, 853-857.
(10) Helton, J. S.; Matan, K.; Shores, M. P.; Nytko, E. A.; Bartlett, B. M.; Yoshida, Y.; Takano, Y.; Suslov, A.; Qiu, Y.; Chung, J. H.; Nocera, D. G.; Lee, Y. S. *Phys. Rev. Lett.* **2007**, *98*, 107204.
(11) de Vries, M. A.; Kamenev, K. V.; Kockelmann, W. A.; Sanchez-Benitez, J.; Harrison, A. *Phys. Rev. Lett.* **2008**, *100*, 157205.
(12) The mineral haydeeite has a similar formula, $MgCu_3(OH)_6Cl_2$, but has a different structure than the compounds reported here.
(13) Hamilton, W. C. *Acta Cryst.* **1965**, *18*, 502.
(14) Zheng, X. G.; Kawae, T.; Kashitani, Y.; Li, C. S.; Tateiwa, N.; Takeda, K.; Yamada, H.; Xu, C. N.; Ren, Y. *Phys. Rev. B* **2005**, *71*, 052409.

Supporting Information

# A Cu$^{2+}$ (S = ½) Kagomé Antiferromagnet: Mg$_x$Cu$_{4-x}$(OH)$_6$Cl$_2$

Shaoyan Chu[‡], Tyrel M. McQueen[†], Robin Chisnell[§], Danna E. Freedman[†], Peter Müller[†],

Young S. Lee[§] and Daniel G. Nocera*,[†]

[†]*Department of Chemistry, [§]Department of Physics, and the [‡]Center for Materials Science and Engineering, Massachusetts Institute of Technology, 77 Massachusetts Avenue, Cambridge, MA 02139 USA*

E-mail: nocera@mit.edu

**Experimental Methods and Preparations**

**General considerations.** Unless otherwise stated, all chemicals were used as received, and all sample manipulations performed in air. Ultrapure water was used (type 1: >18.2 MΩ-cm at 25 ˚C) for sample preparations and pure water (type 2: > 5 MΩ-cm at 25 ˚C) was used for the single crystal growths. Metals analyses were performed at the MIT Center for Materials Science and Engineering Shared Experimental Facility (CSME-SEF) using a HORIBA Jorbin ACTIVA inductively coupled plasma atomic emission spectrometer (ICP-AES). Standards were prepared from materials purchased commercially from Sigma-Aldrich, designated as TraceSELECT grade or better and suitable for ICP analysis. Powder x-ray diffraction patterns were collected on a PANalytical X'Pert Pro instrument using Cu K$_\alpha$ radiation ($\lambda$ = 1.5403 Å) with a Ni foil K$_\beta$ attenuator and a silicon high-speed strip detector. The data were processed and fit using Rietveld techniques with the GSAS program[1] equipped with the EXPGUI interface.[2] Magnetization measurements were performed at the MIT CSME-SEF using a commercial SQUID magnetometer (Quantum Design Magnetic Properties Measurement System, MPMS-5S).

**Preparation of polycrystalline Mg$_x$Cu$_{4-x}$(OH)$_6$Cl$_2$ (1: x = 0.39, 2: x = 0.54, 3: x = 0.75).** A 23 mL teflon liner was charged with 500 mg of Cu$_2$(OH)$_2$CO$_3$ (4.6 mmol Cu), 7 mL of water, and 2.4 g (**1**: 11.6 mmol Mg), 3.1 g (**2**: 15.2 mmol Mg), or 3.8 g (**3**: 18.4 mmol Mg) of MgCl$_2$•6H$_2$O. The liner was capped and placed into a stainless steel pressure vessel (Parr Instrument Company, model #4749). The vessel was heated to 180 ˚C at a rate of 1 ˚C/min, and its temperature was maintained for 48 hr, and then cooled to room temperature at a rate of 0.1 ˚C/min. A blue-green polycrystalline powder was found at the bottom of each vessel, isolated from the liner by filtration, washed with water, and dried over drierite. Rietveld refinement of powder x-ray diffraction of the products confirmed the structure and Mg stoichiometry (see Figs. S1, S2 and S3). ICP-AES analysis found an Mg:Cu ratio of 0.35(3):3.65 for **1**.

**Preparation of single crystals of Mg$_x$Cu$_{4-x}$(OH)$_6$Cl$_2$ (4: x = 0.33, 5: x = 0.65, 6: x = 0.75).** For **4**, 0.20 g of CuO, 2 g of MgCl$_2$•6H$_2$O and 7 mL of H$_2$O were charged in a quartz tube (8 mm ID, 12 mm OD). After purging air by a machine pump, the quartz tube was sealed and pre-reacted in an oven at temperature of 190 ºC for three days. The pre-reaction produced green-blue microcrystalline powder. Millimeter-sized single crystals were grown by the following re-crystallization process: the quartz tube was put upright at room temperature until powder deposited at one end of the tube. Then the tube was laid horizontally in a gradient furnace and slow heated until the hot end, where the powder source material was, reached 190 °C. It was left for 45 weeks, until crystals large enough for single crystal x-ray diffraction and magnetization measurements had formed at the cold end. The temperature gradient at the cool end of the tube,

---

[1] A.C. Larson and R.B. Von Dreele, "General Structure Analysis System (GSAS)", Los Alamos National Laboratory Report LAUR 86-748 (1994).

[2] B. H. Toby, EXPGUI, a graphical user interface for GSAS, *J. Appl. Cryst.* **2001**, *34*, 210-213.

where the crystals nucleated and grew, was measured to be around 1 ºC/cm. Samples **5** and **6** were prepared in a similar fashion but with different starting materials. For both **5** and **6**, 0.125 g (1.93 mmol Mg) of MgO and 0.511 g (3.00 mmol Cu) of $CuCl_2 \cdot 2H_2O$ were used. For **5**, 3.5g $MgCl_2 \cdot 6H_2O$ (17.2 mmol Mg) and 5.4 mL of $H_2O$ were also used, whereas for **6**, 4.5g $MgCl_2 \cdot 6H_2O$ (22.1 mmol Mg) and 5.0 mL of $H_2O$ were added. ICP-AES analysis found an Mg:Cu ratio of 0.24(6):3.74 for **4**.

**Single crystal X-ray crystallographic details.** Low-temperature diffraction data were collected on a Siemens Platform three-circle diffractometer coupled to a Bruker-AXS Smart Apex CCD detector with graphite-monochromated MoKα radiation ($\lambda$ = 0.71073 Å), performing $\varphi$- and $\omega$-scans. Data reduction, including correction for Lorenz and polarization effects was performed SAINT.[3] Absorption correction and scaling, including odd and even ordered spherical harmonics were performed using SADABS.[4] Space group assignments were based upon systematic absences, $E$ statistics, and successful refinement of the structures. The structures were solved with Patterson methods using the program SHELXS[5], and refined against $F^2$ on all data by full-matrix least squares with SHELXL-97[6] using established refinement methods[7]. All non-hydrogen atoms were refined anisotropically. In all structures, coordinates for the hydrogen atom were taken from the difference Fourier synthesis (always corresponding to the highest residual density maximum) and refined semi-freely with the help of a distance restraint, while constraining the $U_{iso}$ of the hydrogen to 1.2 times the value of $U_{eq}$ of the oxygen atom.

For all three structures two sets of refinements were performed. Initially, both the metals in the kagomé plane and the interlayer sites were allowed to refine as a mixture of Cu and Mg. The ratios between two components were refined freely, while the sum of occupancies was constrained to unity. The two atoms overlapping on each metal site were constrained to share coordinates and anisotropic displacement parameters. These refinements were used to generate the tables and cif files. Subsequently the structures were refined without the presence of Mg on the kagomé sites. Details of unit cell parameters, morphology, data quality and a summary of residual values of the refinements for **4-6** are listed in Tables S1, S2, and S3 respectively.

**Magnetization measurements**. The sample holders were plastic straws, measured as having negligible temperature-dependent magnetic response. The temperature-independent diamagnetic signal from the straw was negligible, less than 5% of the smallest raw measured signals for the sample size and fields measured. Core diamagnetism of the sample was corrected by using well-established tables of values for the respective ions present. The magnetic susceptibility was estimated from the magnetization by taking $M / H$, where $M$ is the measured magnetization, and

3  Chambers, J. L. SAINT 7.23, Bruker-AXS, Madison, Wisconsin, USA, 2005.

4  Sheldrick, G. M. SADABS, Bruker AXS, Madison, Wisconsin, USA, 2006.

5  Sheldrick, G. M. *Acta Cryst.* **1990**, *A46*, 467-473.

6  Sheldrick, G. M. *Acta Cryst.* **2008**, *A64*, 112-122.

7  Müller, P. *Crystallography Reviews* **2009**, *15*, 57-83.

*H* is the applied magnetic field. Curie-Weiss parameters were then extracted from fits to the high temperature portion of the data, 150-350 K, using the relation,

$$\frac{1}{\chi} = \frac{1}{C}T - \frac{\theta_{CW}}{C},$$

where *T* is the temperature, *C* is the Curie-Weiss constant, and $\theta_{CW}$ is the Weiss temperature.

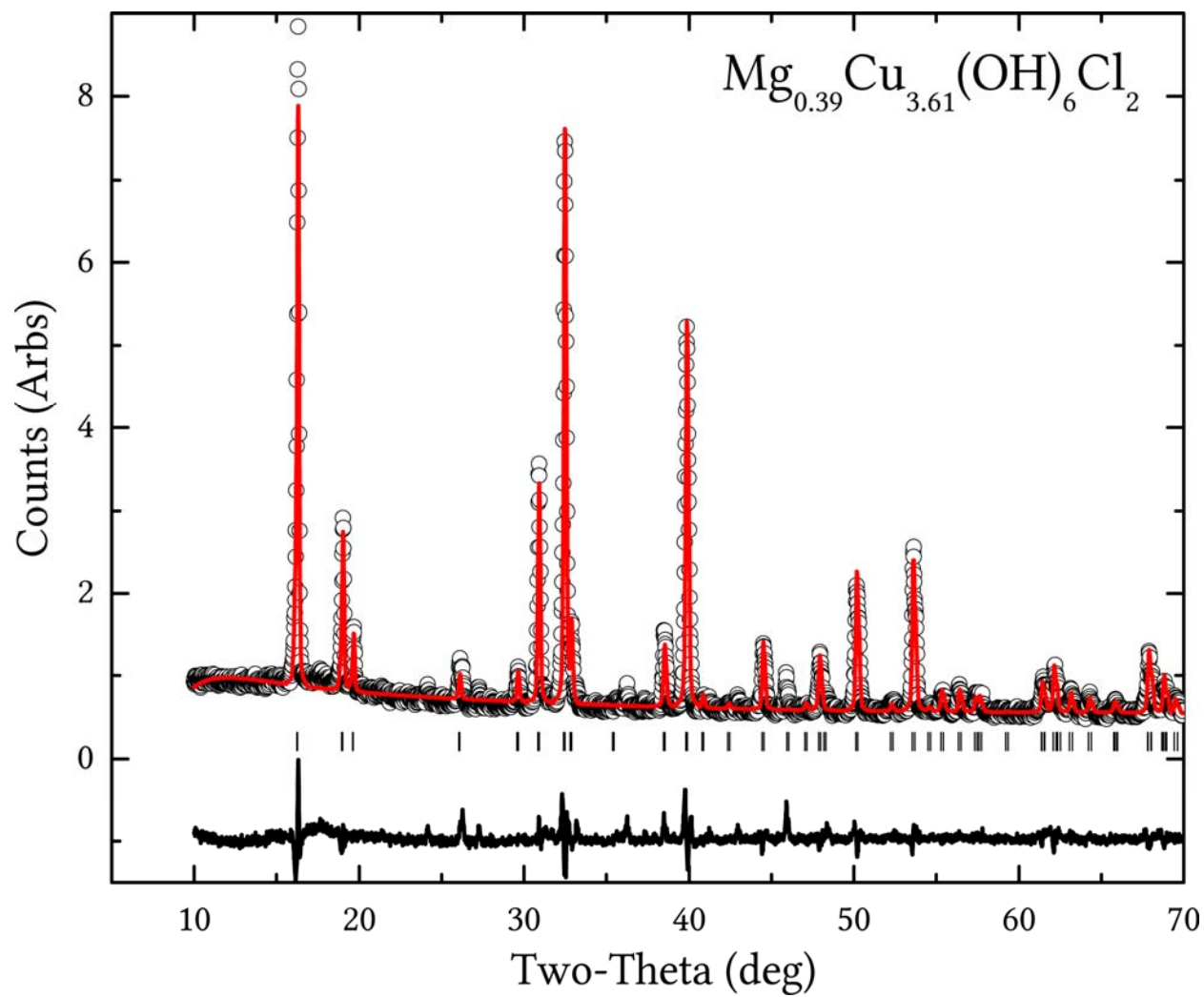

**Figure S1.** Rietveld refinement of powder x-ray diffraction data of **1**. The red line is the fit to the data and the bottom black line is the residual.

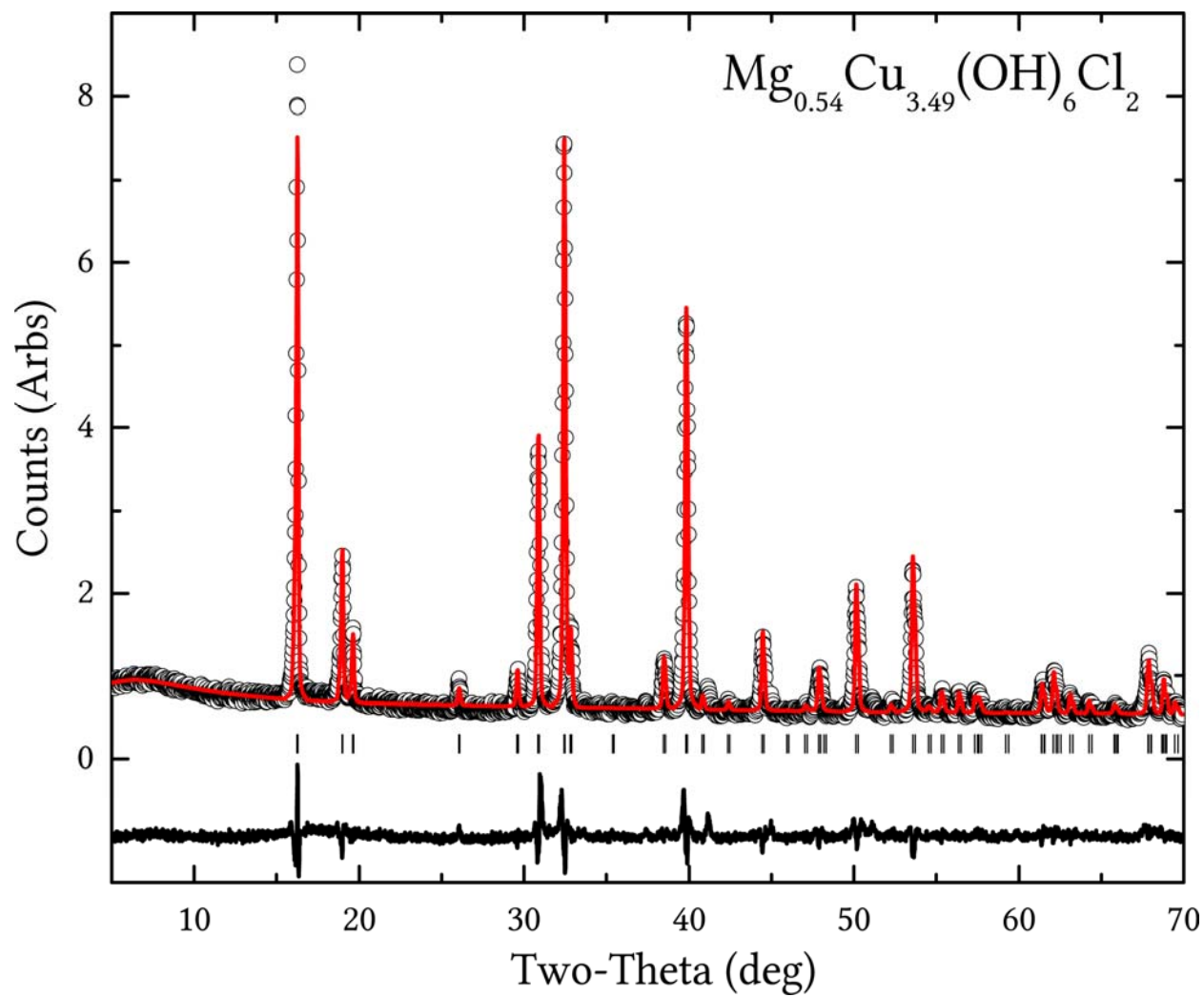

**Figure S2.** Rietveld refinement of powder x-ray diffraction data of **2**. The red line is the fit to the data and the bottom black line is the residual.

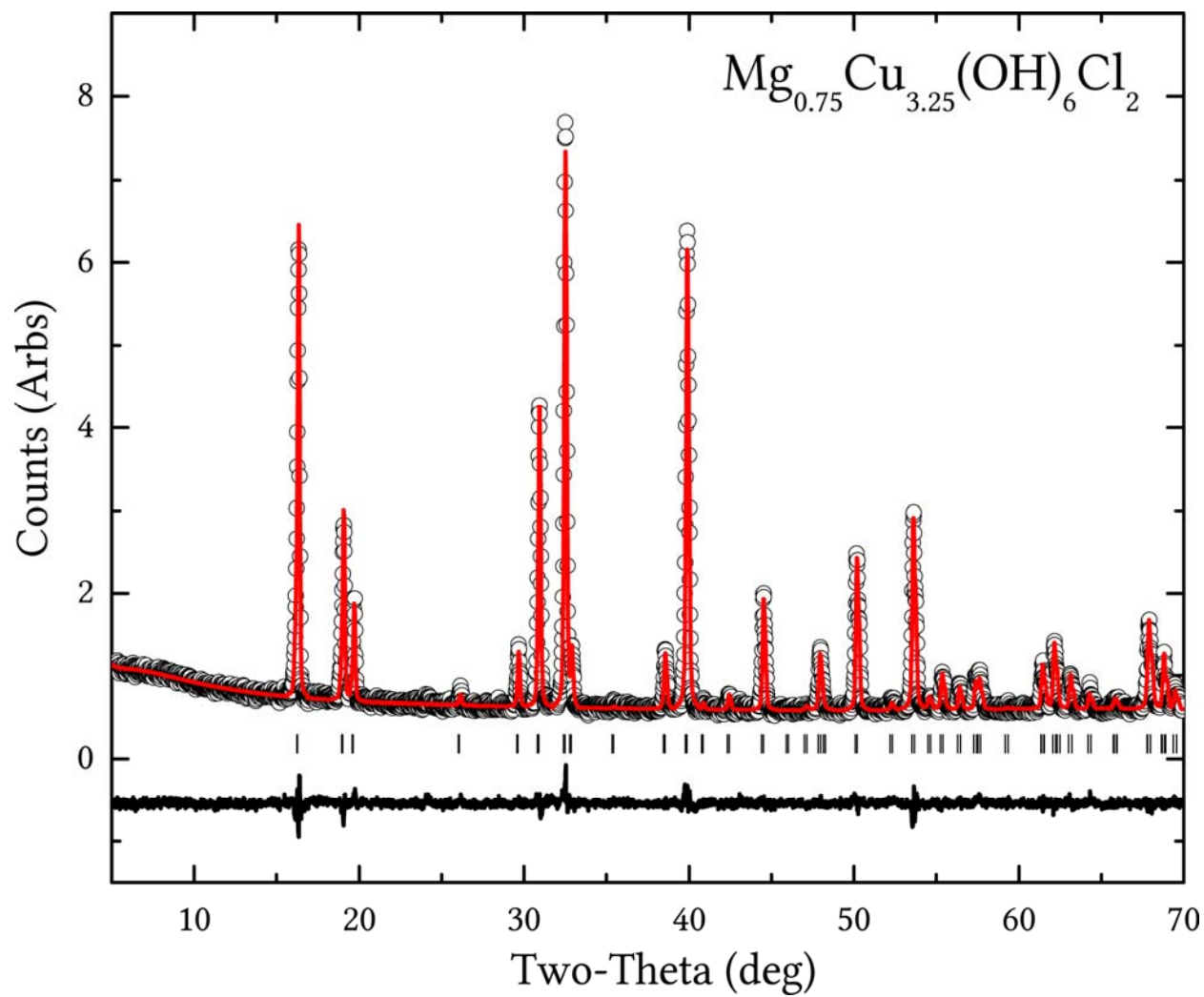

**Figure S3.** Rietveld refinement of powder x-ray diffraction data of **3**. The red line is the fit to the data and the bottom black line is the residual.

**Table S1.** Crystal data and structure refinement for $Cl_2Cu_{3.67}H_6Mg_{0.33}O_6$ (**4**)

| | |
|---|---|
| Empirical formula | $Cl_2Cu_{3.67}H_6Mg_{0.33}O_6$ |
| Formula weight | 414.29 |
| Temperature | −173(2) °C |
| Wavelength | 0.71073 Å |
| Crystal system | Rhombohedral |
| Space group | *R-3m* |
| Unit cell dimensions | $a = 6.8393(18)$Å |
| | $c = 14.005(4)$ Å |
| Volume | 567.3(3) Å$^3$ |
| $Z$ | 3 |
| Density (calculated) | 3.638 mg/m$^3$ |
| Absorption coefficient | 10.913 mm$^{-1}$ |
| F(000) | 595 |
| Crystal size | $0.17 \times 0.16 \times 0.15$ mm$^3$ |
| $\theta$ range for data collection | 3.74 to 29.47° |
| Index ranges | $-9 \leq$ h $\leq 9$, $-9 \leq$ k $\leq 9$, $-19 \leq \ell \leq 19$ |
| Reflections collected | 2585 |
| Independent reflections | 223 [$R_{int} = 0.0414$] |
| Completeness to $\theta = 29.47°$ | 99.1 % |
| Absorption correction | Semi-empirical from equivalents |
| Max. and min. transmission | 0.2914 and 0.2584 |
| Refinement method | Full–matrix least–squares on $F^2$ |
| Data / restraints / parameters | 223 / 2 / 21 |
| Goodness–of–fit on $F^2$ | 1.244 |
| Final R indices [$I>2\sigma(I)$] | $R1 = 0.0219$, $wR2 = 0.0555$ |
| R indices (all data) | $R1 = 0.0221$, $wR2 = 0.0556$ |
| Largest diff. peak and hole | 0.917 and −1.071 e/Å$^{-3}$ |

[a] GOF = $(\Sigma\, w(F_o^2 - F_c^2)^2/(n - p))^{1/2}$ where $n$ is the number of data and $p$ is the number of parameters refined. [b] $R1 = \Sigma||F_o| - |F_c||/\Sigma|F_o|$. [c] $wR2 = (\Sigma(w(F_o^2 - F_c^2)^2)/\Sigma(w(F_o^2)^2))^{1/2}$.

**Table S2.** Crystal data and structure refinement for $Cl_2Cu_{3.35}H_6Mg_{0.65}O_6$ (**5**)

| | |
|---|---|
| Empirical formula | $Cl_2Cu_{3.35}H_6Mg_{0.65}O_6$ |
| Formula weight | 401.74 |
| Temperature | $-173(2)$ °C |
| Wavelength | 0.71073 Å |
| Crystal system | Rhombohedral |
| Space group | *R-3m* |
| Unit cell dimensions | $a = 6.8305(13)$ Å |
| | $c = 13.962(3)$ Å |
| Volume | 564.15(19) Å$^3$ |
| *Z* | 3 |
| Density (calculated) | 3.548 mg/m$^3$ |
| Absorption coefficient | 10.121 mm$^{-1}$ |
| *F*(000) | 579 |
| Crystal size | $0.15 \times 0.13 \times 0.11$ mm$^3$ |
| $\theta$ range for data collection | 3.74 to 30.40° |
| Index ranges | $-9 \leq h \leq 9, -9 \leq k \leq 9, -19 \leq \ell \leq 19$ |
| Reflections collected | 3377 |
| Independent reflections | 239 [$R_{int} = 0.0391$] |
| Completeness to $\theta = 28.34°$ | 100.0 % |
| Absorption correction | Semi-empirical |
| Max. and min. transmission | 0.4023 and 0.3121 |
| Refinement method | Full–matrix least–squares on $F^2$ |
| Data / restraints / parameters | 239 / 1 / 21 |
| Goodness–of–fit on $F^2$ | 1.324 |
| Final *R* indices [$I > 2\sigma(I)$] | $R1 = 0.0168, wR2 = 0.0440$ |
| *R* indices (all data) | $R1 = 0.0175, wR2 = 0.0442$ |
| Largest diff. peak and hole | 0.532 and $-0.454$ e.Å$^{-3}$ |

[a] GOF = $(\Sigma \, w(F_o^2 - F_c^2)^2/(n - p))^{1/2}$ where *n* is the number of data and *p* is the number of parameters refined. [b] $R1 = \Sigma||F_o| - |F_c||/\Sigma|F_o|$. [c] $wR2 = (\Sigma(w(F_o^2 - F_c^2)^2)/\Sigma(w(F_o^2)^2))^{1/2}$.

**Table S3.** Crystal data and structure refinement for $Cl_2Cu_{3.25}H_6Mg_{0.75}O_6$ (**6**)

| | |
|---|---|
| Empirical formula | $Cl_2Cu_{3.25}H_6Mg_{0.75}O_6$ |
| Formula weight | 397.82 |
| Temperature | −173(2) °C |
| Wavelength | 0.71073 Å |
| Crystal system | Rhombohedral |
| Space group | *R-3m* |
| Unit cell dimensions | $a$ = 6.8322(11) Å |
| | $c$ = 13.960(2) Å |
| Volume | 564.33(15) Å$^3$ |
| $Z$ | 3 |
| Density (calculated) | 3.512 mg/m$^3$ |
| Absorption coefficient | 9.851 mm$^{-1}$ |
| $F(000)$ | 574 |
| Crystal size | 0.17 × 0.16 × 0.15 mm$^3$ |
| $\theta$ range for data collection | 3.74 to 29.89° |
| Index ranges | −8 ≤ h ≤ 9, −9 ≤ k ≤ 9, −19 ≤ ℓ ≤ 19 |
| Reflections collected | 2792 |
| Independent reflections | 228 [$R_{int}$ = 0.0340] |
| Completeness to $\theta$ = 29.89° | 99.6 % |
| Absorption correction | Semi-empirical from equivalents |
| Max. and min. transmission | 0.3196 and 0.2852 |
| Refinement method | Full−matrix least−squares on $F^2$ |
| Data / restraints / parameters | 228 / 1 / 21 |
| Goodness−of−fit on $F^2$ | 1.194 |
| Final $R$ indices [$I>2\sigma(I)$] | $R1$ = 0.0162, $wR2$ = 0.0414 |
| $R$ indices (all data) | $R1$ = 0.0172, $wR2$ = 0.0421 |
| Largest diff. peak and hole | 0.464 and −0.729 e.Å$^{-3}$ |

[a] GOF = $(\Sigma\, w(F_o^2 - F_c^2)^2/(n-p))^{1/2}$ where $n$ is the number of data and $p$ is the number of parameters refined. [b] $R1 = \Sigma||F_o| - |F_c||/\Sigma|F_o|$. [c] $wR2 = (\Sigma(w(F_o^2 - F_c^2)^2)/\Sigma(w(F_o^2)^2))^{1/2}$.

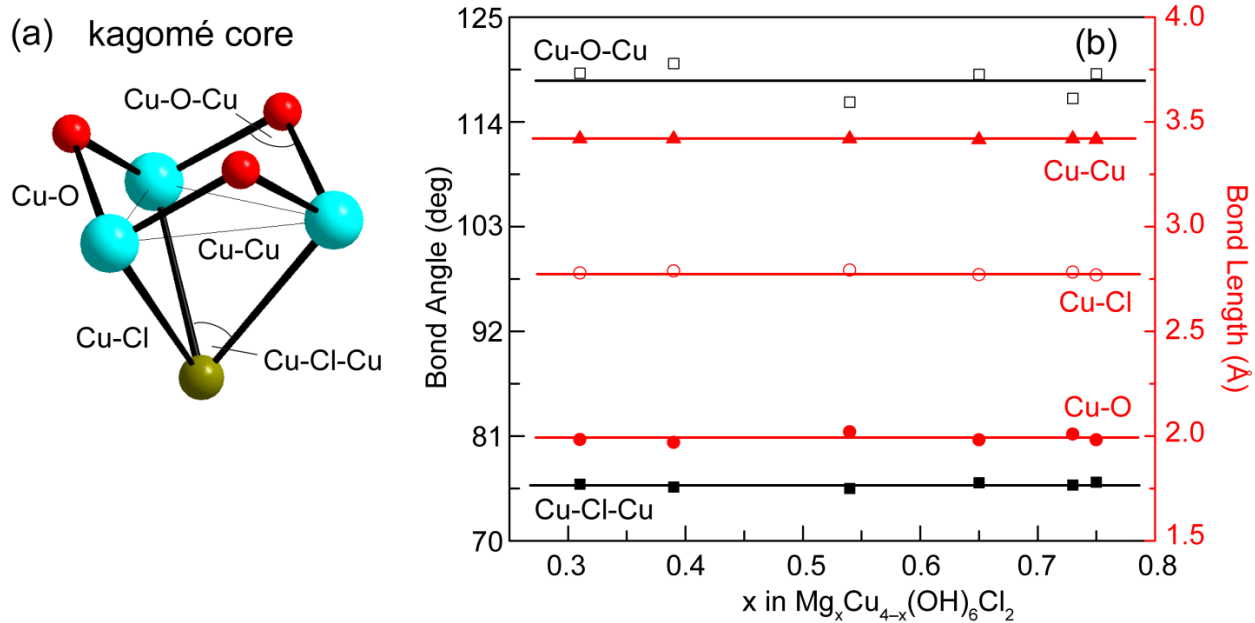

**Figure S4.** (a) The core structural motif of the kagomé layers in $Mg_xCu_{4-x}(OH)_6Cl_2$ (Cu: blue, O: red, Cl: mustard), with unique bond distances and angles identified. (b) Structural parameters of the core unit, from single crystal and powder x-ray diffraction refinements. There are no systematic changes as a function of magnesium content.

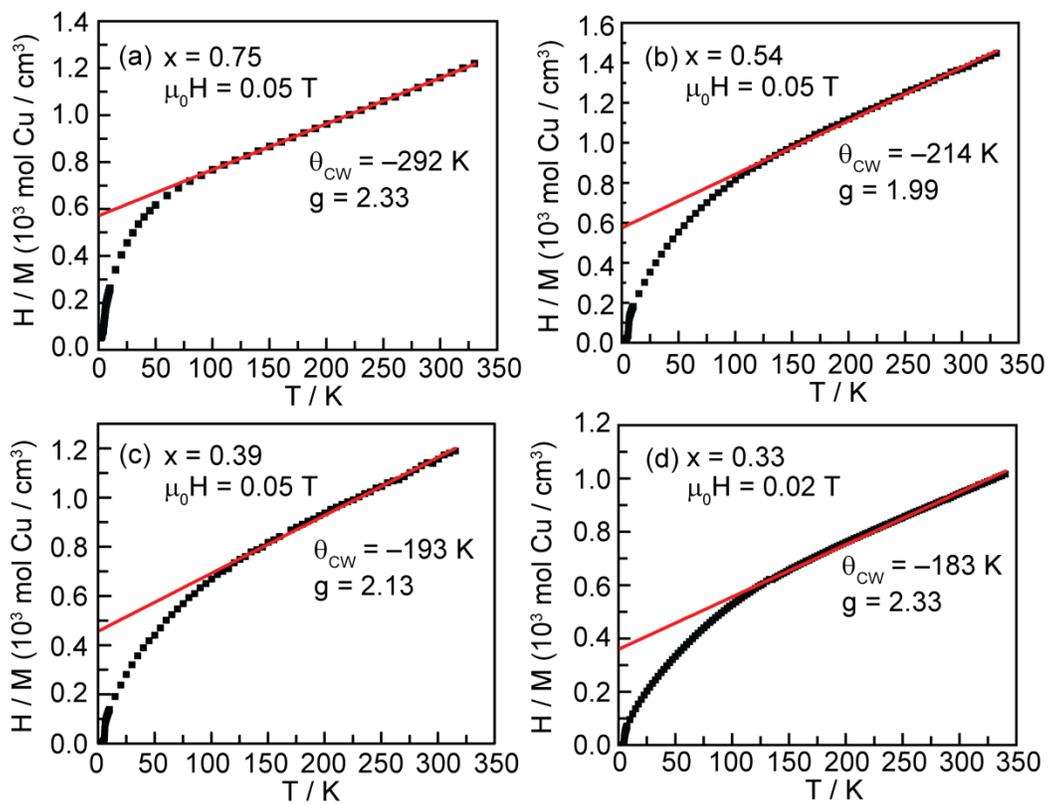

**Figure S5.** Curie-Weiss fits of the inverse magnetic susceptibility of Mg$_x$Cu$_{4-x}$(OH)$_6$Cl$_2$ for (a) **3**, x = 0.75, (b) **2**, x = 0.54, (c) **1**, x = 0.39, and (d) **4**, x = 0.33.

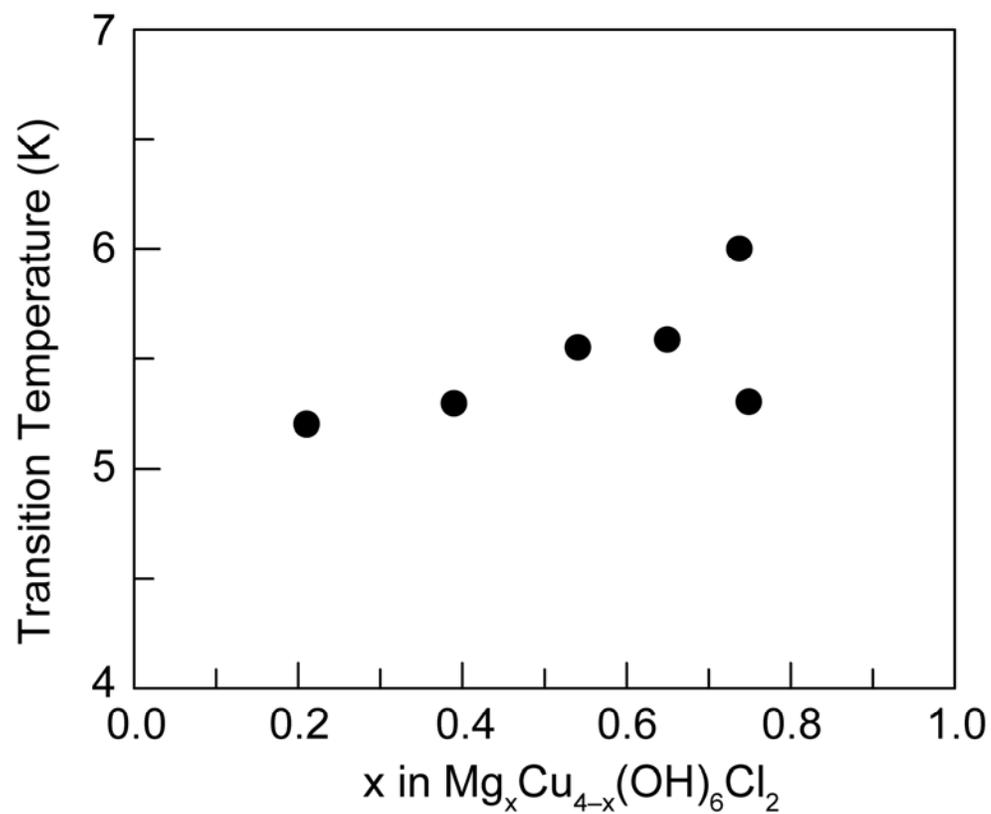

**Figure S6.** The temperature at which the magnetic transition occurs, determined from the data presented in manuscript Figure 2 as the temperature where M/H significantly deviates from the high temperature trend.